\documentclass[twoside,a4paper,11pt]{torus2012}
\usepackage{graphicx}
\usepackage{hyperref}
\usepackage{movie15}

\begin{document}
\pagenumbering{arabic}
\pagestyle{myheadings}
\thispagestyle{empty}
{\flushleft\includegraphics[width=\textwidth,bb=58 650 590 680]{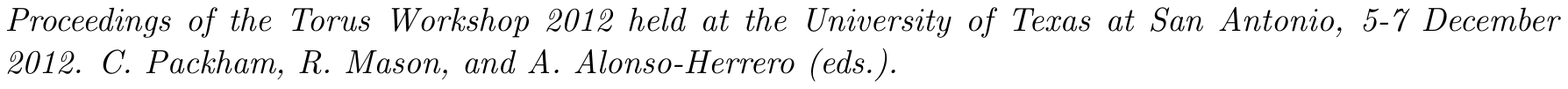}}
\vspace*{0.2cm}
\begin{flushleft}
{\bf {\LARGE
%
Why a Windy Torus?
%
}\\
\vspace*{1cm}
%
S. C. Gallagher$^{1}$,
M. M. Abado$^{1}$,
J. E. Everett$^{2}$, 
S. Keating$^{3}$,
and 
R. P. Deo$^{4}$
}\\
\vspace*{0.5cm}
%
$^{1}$
University of Western Ontario\\
$^{2}$
Northwestern University\\
$^{3}$
University of Toronto\\
%
\end{flushleft}
%
\markboth{
Why a Windy Torus?
}{ 
%
Gallagher et~al.
%
}
\thispagestyle{empty}
\vspace*{0.4cm}
\begin{minipage}[l]{0.09\textwidth}
\ 
\end{minipage}
\begin{minipage}[r]{0.9\textwidth}
\vspace{1cm}
\section*{Abstract}{\small
%

Mass ejection in the form of winds or jets appears to be as
fundamental to quasar activity as accretion, and can be directly
observed in many objects with broadened and blue-shifted UV absorption
features.  A convincing argument for radiation pressure driving this
ionized outflow can be made within the dust sublimation
radius. Beyond, radiation pressure is even more important, as high energy
photons from the central engine can now push on dust grains.  This
physics underlies the dusty-wind model for the putative obscuring
torus.  Specifically, the dusty wind in our model is first launched
from the outer accretion disk as a magneto-centrifugal wind and then
accelerated and shaped by radiation pressure from the central
continuum.  Such a wind can plausibly account for both the
necessary obscuring medium to explain the ratio of
broad-to-narrow-line objects and the mid-infrared emission commonly seen in
quasar spectral energy distributions.  A convincing demonstration that
large-scale, organized magnetic fields are present in radio-quiet
active galactic nuclei is now required to bolster the case for this
paradigm.

%
\normalsize}
\end{minipage}
%
%
%
\section{Introduction \label{intro}}

Researchers currently studying active galactic nuclei agree on a few
key premises.  The central engine is powered by a supermassive black
hole, and the optical through ultraviolet continuum emission is
generated by optically thick material in an accretion disk within the
central parsec down to a few gravitational radii.  Important specifics
such as the thickness of the disk as a function of radius are still
being hashed out, but the overall picture is consistent; it is hard to
avoid disk formation given the persistence of angular momentum.
Moving to larger radii, ionized gas moving at high velocities covers
approximately $\sim10$\% of the sky and gives rise to broad
(1000s~km~s$^{-1}$) resonance and semi-forbidden emission lines that
characterize Type~1 quasars and Seyfert galaxies.  In many luminous
quasars, the broad UV emission lines show signatures of winds.  The
most obvious are the P Cygni profiles seen in the high ionization
resonance lines of Broad Absorption Line (BAL) quasars (e.g.,
\cite{weymann+91}).  More subtle are the blue-shifts and asymmetries
seen in the same lines in other luminous quasars (e.g.,
\cite{richards+11}).  The wind manifest in these lines is understood
to be driven by radiation pressure on ions at UV resonance
transitions.  For the wind to be launched in the vicinity of the
broad-line region, the quasar continuum must not only have significant
power in the UV -- the source of the line-driving photons -- but not
too much power in the X-ray.  X-rays will strip atoms such as C and O
of their electrons, thus dropping the absorption cross section
dramatically.  Overionized gas can therefore only be driven by Thomson
scattering, which is much less efficient than line driving
(e.g., \cite{everett_ballantyne04}).  The elegant picture of Murray et
al. \cite{murray+95} that unifies the broad emission lines seen in all
quasars with the BAL wind directly observed in only $\sim20$\% (e.g.,
\cite{hewett_foltz03}) by generating broad emission lines in an
equatorial disk wind is compelling, but has not been widely adopted in
other models. Broad-line ``clouds'' are still evoked as the source of
the broad-line emission, though the smoothness of the broad-line
profiles and the requirement of some means of pressure-confinement to
prevent the putative clouds from shredding remain persistently
unsolved problems in the cloud picture (e.g., \cite{murray_chiang97}).

One structure that is even more uncertain is the so-called torus.  The
inner radius of the torus is set by the dust sublimation radius,
beyond which some fraction of the accretion disk continuum is absorbed
by grains and reradiated.  Operationally, the torus must account for
the $\sim30\%$ of the integrated radiant quasar power that comes out
in the near-to-mid-infrared.  Furthermore, it must obscure the central
continuum and broad-line region in a significant fraction of Type~2
objects.  While a static toroidal structure (whether smooth or clumpy)
beyond the dust sublimation radius will serve these purposes, such a
cold (100--1500~K) structure with the dust mass implied by the
mid-infrared luminosity would quickly become gravitationally unstable
and collapse.  This is an old problem, which was addressed by K\"onigl
\& Kartje \cite{konigl_kartje94} with a magneto-hydrodynamically
launched wind, shaped also by radiation pressure, on dust.  This
picture was expanded by our group (\cite{keating+12}; hereafter K12)
by incorporating dust-radiation driving into the magneto-hydrodynamic
(MHD) wind models of Everett \cite{everett05}, and then illuminating the wind
with the central continuum to generate a model spectral energy
distribution (SED).  While the details of the shape of the output
continuum still require some refining to match those observed, the
overall power and general shape generated by our `fiducial' model is
promising.  This model uses as inputs the empirical SDSS composite
quasar continuum \cite{richards+06}, a black hole mass of $M_{\rm
  BH}=10^{8}$~$M_\odot$, an Eddington ratio of $L/L_{\rm Edd}=0.1$,
and a column density at the base of the wind of $N_{\rm
  H,0}=10^{25}$~cm$^{-2}$.  We assume a Milky Way interstellar-medium
dust distribution.  In this article, we expand on the results
presented in K12 by exploring how key input model properties affect
the structure of the wind.

\section{The Dusty Wind Model}

Following Blandford \& Payne \cite{blandford_payne82} and K\"onigl \&
Kartje \cite{konigl_kartje94}, we model the torus as a self-similar
dusty wind driven by MHD forces and radiation pressure. The MHD
radiative wind code has been described in more detail previously \cite{everett05,keating+12}, and we give only a brief outline here. First, a
purely MHD-driven wind solution is calculated given a set of initial
conditions input by the user.  Next, Cloudy (version 06.02.09b, as
described by \cite{Cloudy}) is called to calculate the radiation field
and dust opacity in the wind, allowing for the model to determine the
radiative acceleration of the wind due to the central accretion disk
continuum.  A new MHD solution is calculated for the wind, now taking
into account the radiative acceleration, and the process is iterated
until the wind converges to a stable profile.

Operationally, our assumption of self-similarity means that several
important quantities, including magnetic field strength and mass
density, scale with spherical radius; this simplifies the calculation
of the solution to the MHD equations considerably.  Furthermore, it
allows for the incorporation of radiation pressure from the disk by
simply reducing the effective gravitational potential. Importantly, it
allows us to use one streamline as an approximate template for the entire wind,
simply scaling appropriately inwards or outwards to find properties of
the specific region of the wind under consideration.

The final, converged MHD radiative wind model therefore provides the
wind structure necessary to calculate the mass density of the wind as
a function of \emph{r}, the radial distance from the axis of rotation
of the disk, and \emph{z}, the vertical distance from the plane of the
disk. The model supplies the density along each streamline as a
function of the mass density at the base of the streamline and of
polar angle. The density of the wind at any arbitrary point is then
relatively easy to calculate as a consequence of the self-similarity
assumption. Because we know the shape of the wind streamlines, we can
use the shape of a single streamline to trace back from an arbitrary
point to the disk-launching radius for any wind streamline.  From
there, we use our knowledge of the number density of particles on the
disk at the inner launch radius and the fact that it is assumed to be
proportional to $r^{-1.5}$ \cite{blandford_payne82} to find the
number density at the launch radius of that particular streamline. We
convert from mass density to number density by assuming that the
average particle has the mass of a proton, thus allowing us to find
the number density at our specified point.

\section{Properties of the Wind}

\begin{figure}
\center
\includegraphics[width=8.5cm]{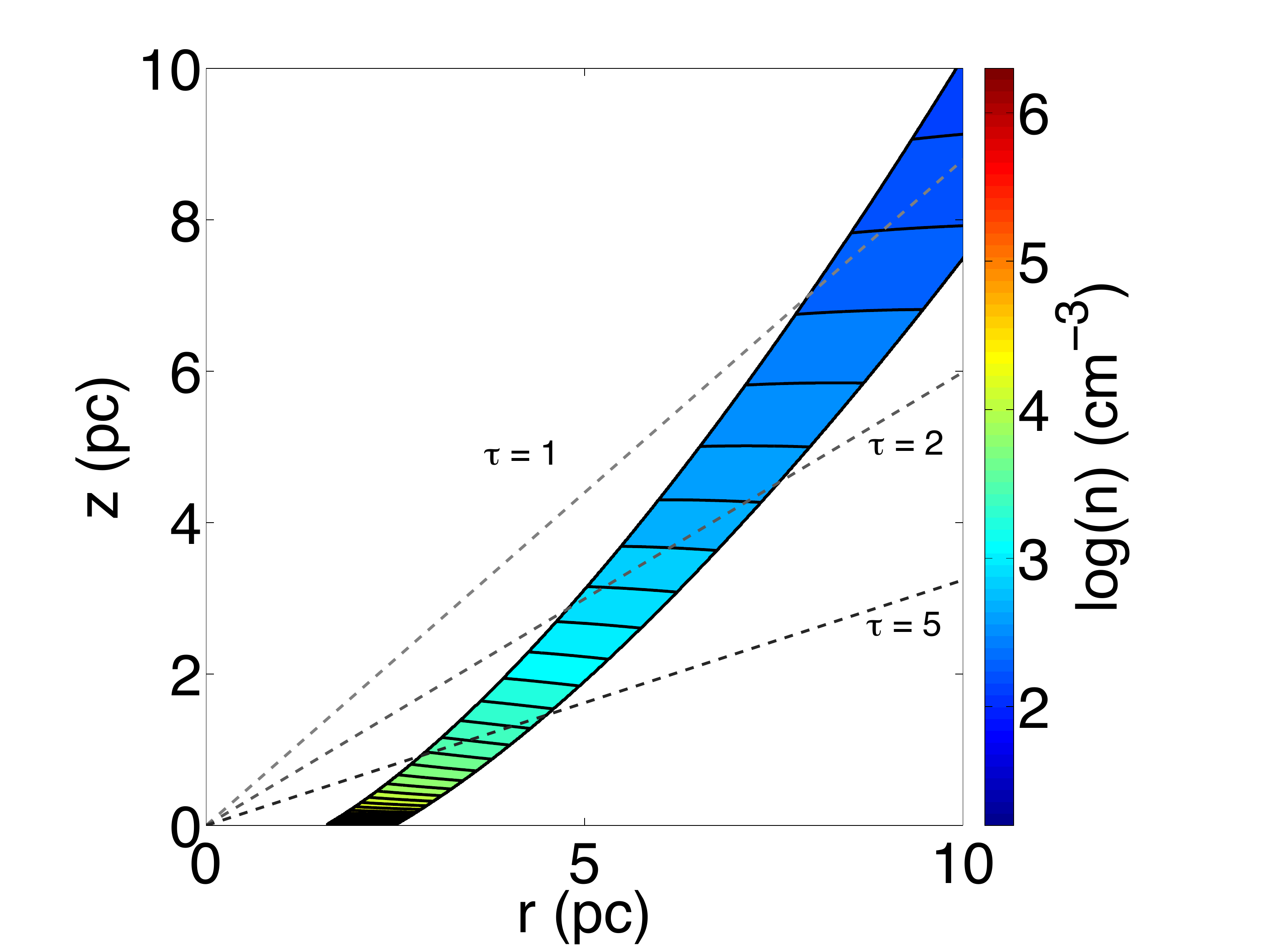} ~
\raisebox{18pt}{\includegraphics[width=4.8cm]{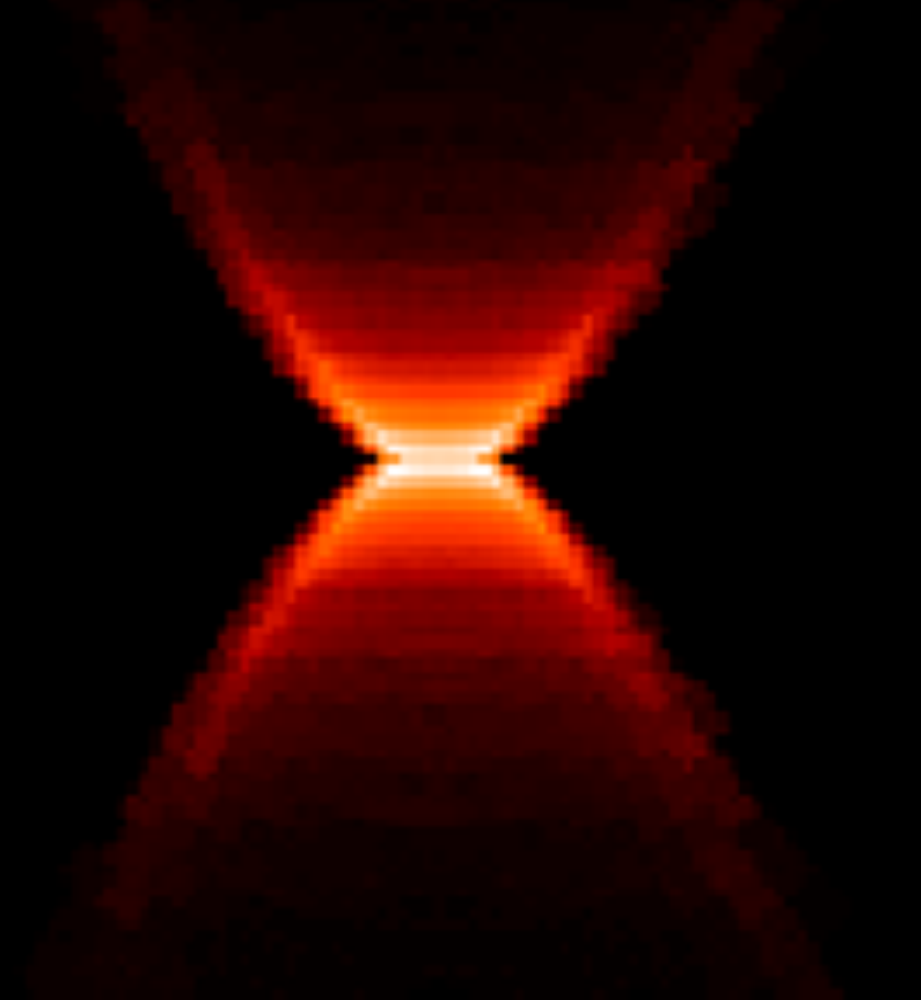}} ~
\caption{\label{fig1} {\em Left:} A 2D cross-section of the upper right quadrant of the number density map of the fiducial dusty wind model with $N_{\rm
H,0}=10^{25}$~cm$^{-2}$, $M_{\rm BH}=10^8~M_\odot$, and $L/L_{\rm
Edd}=0.1$.  The color indicates number density according to the scale bar on the right. The number density drops sharply as a function of height,
$z$, and the wind is quite narrow.  Dotted straight lines indicate the values of the optical depth $\tau$ at H$\alpha\lambda6563$, as labeled.   {\em Right:} A simulated high-resolution
9.5~$\mu$m image of the same model generated by MC3D \cite{wolf03}.
The bulk of the emission comes from the inner streamlines heated by
the accretion disk continuum.  }
\end{figure}

In Figure~\ref{fig1} we present a color map plot of the number
density of the dusty wind as a function of \emph{r} and \emph{z} for
the fiducial model. The inner wind launch radius is set by the dust
sublimation radius. The notable thinness of the wind is a
consequence of the model parameters: in general, the outer wind launch
radius is given by:
\begin{equation}
  r_{\rm outer} = r_{\rm in}\left(1-\frac{N_{\rm H, 0}}{2n_{\rm in}r_{\rm in}}\right)^{-2}
\label{eq: thickness}
\end{equation}
for user-specified values of the inner launch radius $r_{\rm in}$, the
number density at the inner launch radius $n_{\rm in}$, and the column
density along the base of the wind $N_{\rm H, 0}$.  A simulated high-resolution image
of the wind at 9.5~$\mu$m heated by the central continuum is shown in
the right panel of Fig.~\ref{fig1}.  The hourglass shape is
characteristic of MHD winds, and reminiscent of the outflows seen in
young stars. It illustrates clearly that the base of the wind and the
inner streamlines dominate the emission at this wavelength; the
brightest 9.5~$\mu$m emission regions would correspond to the largest
extinction values in the optical.  Eight to 13~$\mu$m interferometric
observations have led H\"onig et al. \cite{hoenig+12} to infer a similar structure
and characteristic size scale ($\sim2$~pc) for the dust distribution
in the Seyfert~2 galaxy NGC~424.

\begin{figure}
\center
\includegraphics[width=10.0cm]{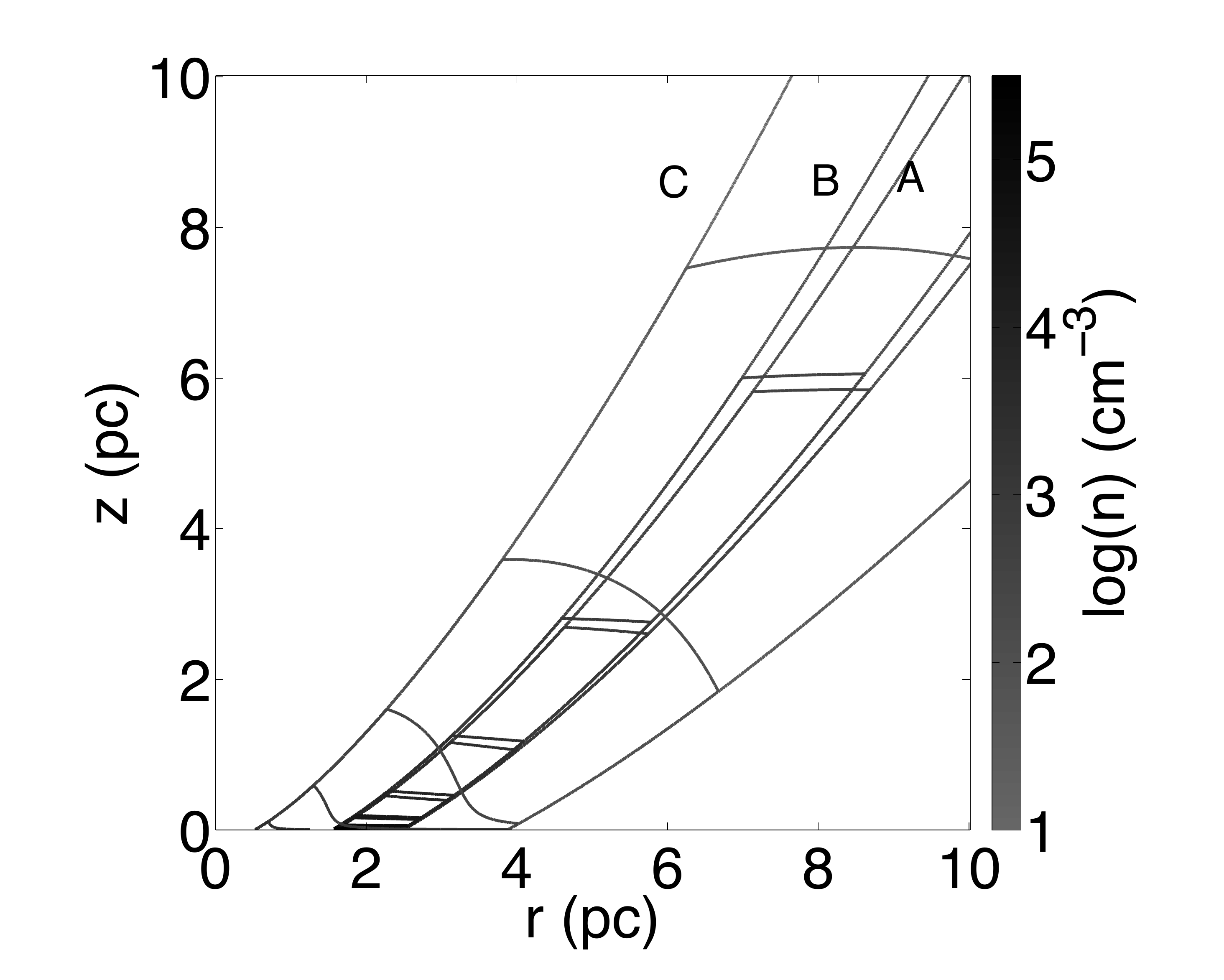} ~
\caption{\label{fig2} A cross section of the wind, as in Figure~\ref{fig1},
  comparing the shape of several wind models.  Shown here is the
  fiducial model (labelled ``A''), with $N_{\rm
    H,0}=10^{25}~\rm{cm}^{-2}$, $L/L_{\rm Edd}=0.1$, and $M_{\rm
    BH}=10^8~\rm{M}_{\odot}$; a similar
  model with $M_{\rm BH}=10^9~\rm{M}_{\odot}$ (``B''); and a third model
  with the same parameters as the fiducial model, but with a smaller
  inner launch radius (``C''). Increasing the black hole mass by a
  factor of 10 does not significantly change the shape of the wind,
  but reducing the inner launch radius from $4.85 \times 10^{18}$~cm
  to $1.57 \times 10^{18}$~cm while maintaining the same density at
  the inner launch radius causes the wind to widen dramatically, a
  consequence of our assumptions.  The movement to smaller radii in model ``B'' also results in a strong near-infrared bump to appear in the output SED \cite{keating+12}.
}
\end{figure}

\begin{figure}
\center
\includegraphics[width=13.0cm]{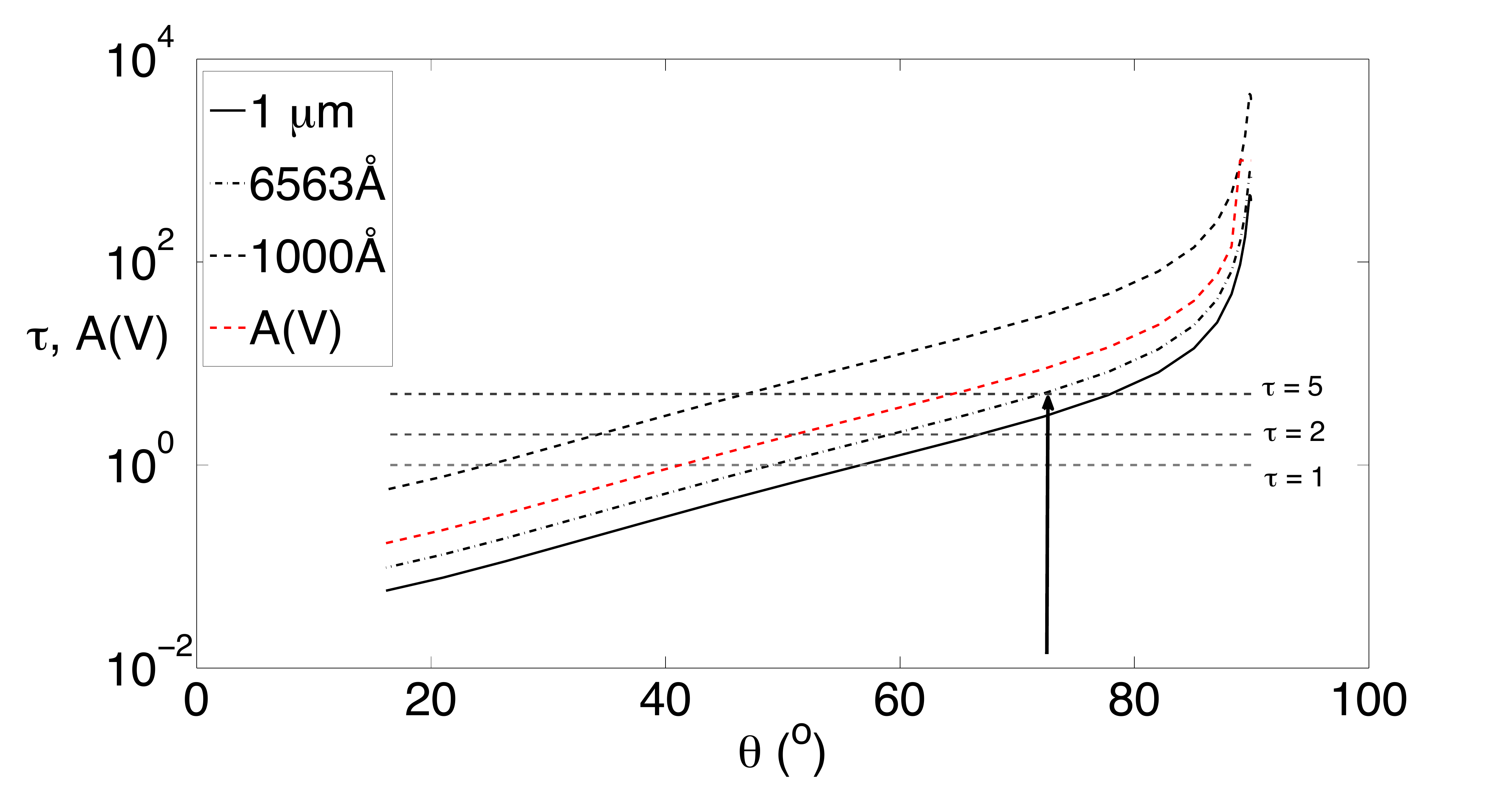} ~
\caption{\label{fig3} The optical depth at three characteristic wavelengths (1~$\mu$m, 6563~\AA, and 1000~\AA) through the dusty wind plotted as a
  function of inclination angle. Shown here is the fiducial model,
  with $N_{\rm H,0}=10^{25}~\rm{cm}^{-2}$, $L/L_{\rm Edd}=0.1$, and
  $M_{\rm BH}=10^8~\rm{M}_{\odot}$. The grey dashed lines
  indicate $\tau$ values of 1, 2, and 5 as labeled. The
  intersection
  of the curve corresponding to 6563~\AA\ and the line corresponding
  to $\tau=5$ determines the
 opening angle (marked with a vertical
  black line). The red dashed curve just above the optical depth curve at
  6563~\AA\ is the $V$-band extinction; $A_{V}$ has a value of 7.7 for $\tau_{\rm H\alpha}=5$.}
\end{figure}

As luminous quasars are typically farther away and therefore fainter
and of smaller angular size compared to nearby Seyferts, they are
beyond the reach of the current generation of interferometers.
Nonetheless, within the context of our model we can explore what might
be expected by investigating the effects of changing luminosity,
$L/L_{\rm Edd}$, and $M_{\rm BH}$ on the structure of the wind.
Figure~\ref{fig2} illustrates the shape of the fiducial wind model
(labeled as A) compared to two other models. Model B has a black hole
mass of $10^9~M_{\odot}$, ten times that of the fiducial model.
Because the value of $L/L_{\rm Edd}$ is fixed at 0.1, this model is 10
times more luminous than the fiducial model.  The resultant shape of
the wind is nearly identical to that of the fiducial model; the
streamlines are slightly more vertical. The third model, C, is also
similar to the fiducial model, this time changing only the inner wind
launch radius to approximately one third of the fiducial model's inner
radius.  The change in the inner radius was made to approximate the
difference between the launch radius of graphites (with a higher
sublimation temperature) versus silicates. This decrease in the inner
launch radius causes the wind to be notably broader in our model (see
Fig.~\ref{fig3}), as expected from Equation~\ref{eq: thickness}.
Because the bulk of the near-infrared emission is coming from the innermost
streamlines, it therefore follows that a smaller $r_{\rm launch}$ for
a given $L/L_{\rm Edd}$ generates more near-IR emission because of the
higher dust temperatures \cite{keating+12}.  The strength of the 3--5~$\mu$m bump
-- the manifestation of a larger contribution to the near-to-mid-infrared
SED from hot dust -- is seen empirically to increase with increasing
quasar luminosity (e.g., \cite{richards+06,gall+07,coleman+13}).
Within the dusty wind paradigm, this implies some mechanism for
decreasing the wind launching radius as a function of luminosity which is challenging to arrange.  A
second way of reducing $r_{\rm launch}$ is to reduce $L/L_{\rm Edd}$
while keeping the black hole mass constant.  This is
also consistent with the expected movement of the dust sublimation boundary to
smaller radii.  The slight change in wind structure from altering
$M_{\rm BH}$ but keeping $L/L_{\rm Edd}$ constant compared with the
large change from decreasing $L/L_{\rm Edd}$ by a factor of 10 shows
that the Eddington ratio is the relevant driver given our assumed wind
structure. However, while the
effects on the wind structure are noticeable when compared to the
fiducial model, this physical difference did not lead to a significant change in the shape
of the output SED \cite{keating+12}.


The column densities of 30 elements at each ionization state as well
as those of several important molecules are calculated by Cloudy.
Because they are part of a wind, these column densities are smooth
functions of inclination angle, monotonically increasing toward the
plane of the disk, where the polar angle $\theta=90^\circ$. Therefore,
they offer no obvious inclination angle to choose as the boundary for
our wind. Therefore, a plot of column density versus height does not
show any obvious features that define a `height' for the wind.  Also
available from Cloudy is the cross section per proton for the
dust in the wind as a function of photon energy. When multiplied by
the total column density of hydrogen (i.e., both neutral and ionized),
this serves as a measure of the optical depth of the
wind. Figure~\ref{fig3} shows the optical depth as a function of polar
angle $\theta$ for a few selected wavelengths. This optical depth is also
 a smooth function of $\theta$ for a wide range of
wavelengths, and similarly provides no obvious angle to mark the
boundary of the wind.  We have therefore chosen the angle at which the
dust reaches an optical depth of 5 for light at a wavelength of
6563~\AA\ (the wavelength of $H\alpha$ emission) to be a reasonable
definition of the opening angle of the dusty wind. At this value of
$\tau$, only 0.7\% of the broad $H\alpha$ emission is transmitted
through the wind, and so a quasar viewed from this inclination angle
(or closer to the disk) would be classified as an optical narrow-line object.

\begin{figure}
\center
\includegraphics[width=13.0cm]{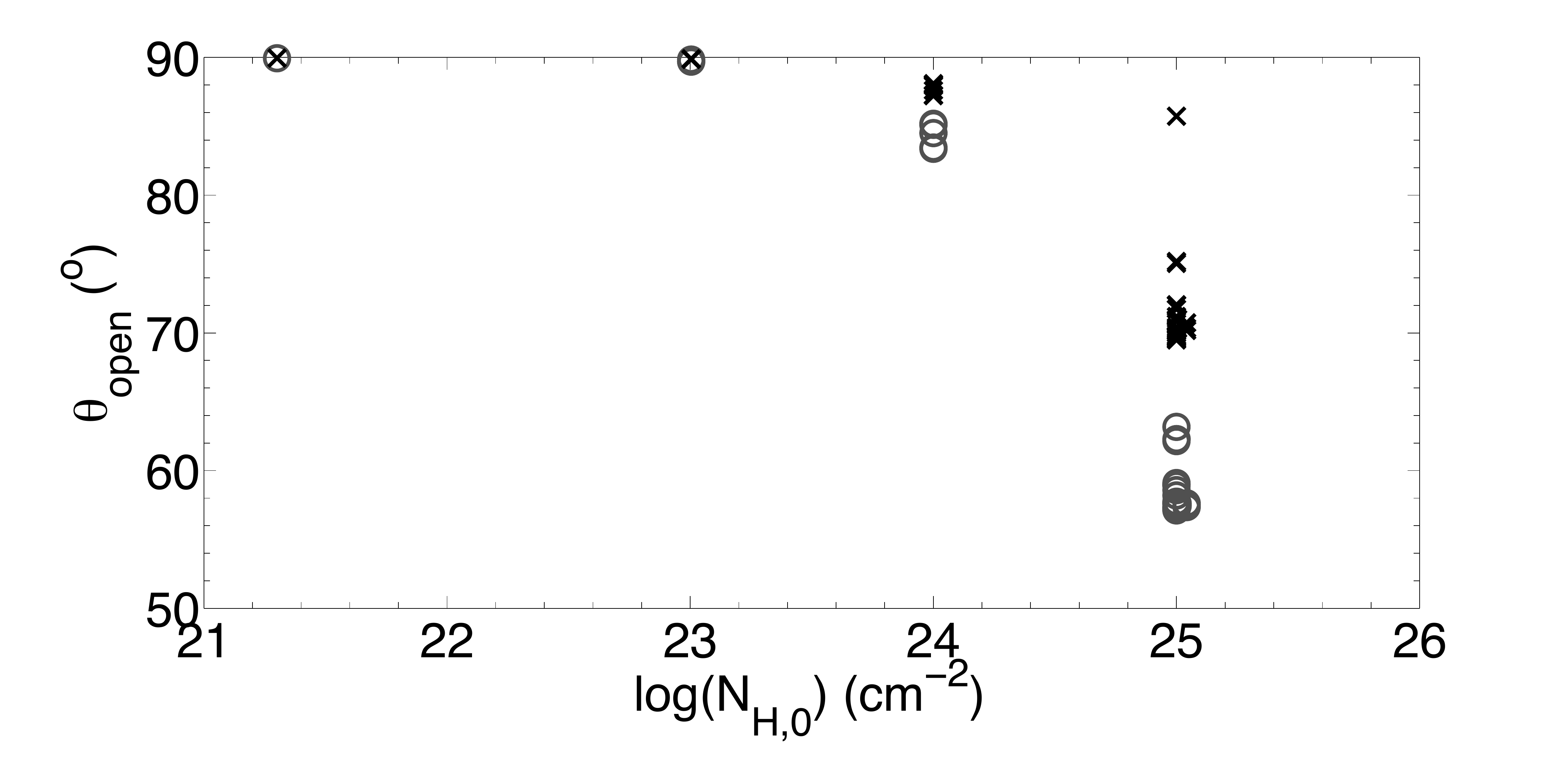} ~
\caption{\label{fig4} The opening angle, $\theta_{\rm open}$, as a
function of base column density, $N_{\rm H,0}$.  At small base column
densities, the opening angle is implausibly large.  For $N_{\rm
H,0}=10^{25}$~cm$^{-2}$, the values drop to $\sim70^\circ$.  This is
also the value of $N_{\rm H,0}$ that generates appropriate
near-to-mid-infrared power in the SED \cite{keating+12}.  The black crosses denote the
opening angle as defined in the text: the angle at which the wind
reaches an optical depth of 5 to the broad H$\alpha$ emission
line. The grey circles show the angle at which the same models reach
an optical depth of 2 to broad H$\alpha$ emission.  The models are
those presented in K12.
}
  \end{figure}

Typical opening angles using this criterion are around 70$^\circ$,
which would lead to Type~2 to Type~1 ratios of $\sim20-25$\%.  This is
on the low end of the range of estimates (e.g., \cite{hao+05,
simpson05}).  However, if accretion disks are randomly oriented with
respect to their host galaxies as expected, the observed Type~2
fractions are inflated with respect to the torus-covering fraction
because of line-of-sight obscuration occurring from dust within the
host galaxy.  The modeled range of opening angles is tight, and shows
little dependence on model input parameters with the exception of the
column density at the base of the wind, and the inner wind launch
radius. This can be seen in Figure~\ref{fig4}, where there is a clear
trend relating the opening angle and the base column density with
relatively little spread among the suite of models from K12.

\section{Conclusions and Future Work}

Getting the dusty-wind model to the point where it can plausibly
account for the detailed structure in near-to-mid-infrared quasar SEDs
requires additional refinement.  In particular, we have not taken into
account the expected dust-free component of the wind that is interior
to the dust sublimation radius.  The strong absorption within the wind
as seen in the X-ray and UV spectra of BAL quasars will significantly
modify the continuum impinging upon the dusty wind.  Specifically, we
expect the dust sublimation radius to be pushed to smaller values.
However, the specific effects of the broad-line region gas shielding
the dusty wind from the continuum will depend on its column density
and ionization state; this is the subject of ongoing investigations.

More generally, the viability of the dusty MHD disk wind paradigm
requires an investigation into whether relatively strong (mG to G),
large-scale poloidal magnetic fields persist within the central
parsecs of quasar host galaxies.  Certainly quasars with jets have
magnetic fields; radio synchrotron emission is observed.  Galaxies
also show organized magnetic fields on large scales, but they are much
weaker, of order 10~$\mu$G (e.g., \cite{beck+96}).  Within the dense
cores of molecular clouds in our Galaxy, magnetic fields with
strengths up to a few mG have been measured
(e.g., \cite{crutcher99}). The physical scale and $H_2$ number
density of such cores (0.02--20~pc and $10^{3-7}$~cm$^{-3}$) span that
expected in tori.  In the near future, the available sensitivity and
spatial resolution of {\it ALMA} and the Jansky VLA are promising for
placing meaningful constraints on the strengths of organized fields
within the central few parsecs of radio-quiet active galactic nuclei.
%
%
\small  
\section*{Acknowledgments}   
%
This work was supported by the Natural Science and Engineering
Research Council of Canada and the Ontario Early Researcher Award
Program.



\begin{thebibliography}{}
\small
%
\bibitem{beck+96}{{Beck}, R., {Brandenburg}, A., {Moss}, D., {Shukurov}, A., \& {Sokoloff}, D. 1996, ARAA, 34, 155}
\bibitem{blandford_payne82}{{Blandford}, R.~D. \& {Payne}, D.~G. 1982, MNRAS, 199, 883}
\bibitem{crutcher99}{{Crutcher}, R.~M. 1999, ApJ, 520, 706}
\bibitem{everett05}{{Everett}, J.~E. 2005, ApJ, 631, 689}
\bibitem{everett_ballantyne04}{{Everett}, J.~E. \& {Ballantyne}, D.~R. 2004, ApJ, 615, L13}
\bibitem{Cloudy}{{Ferland}, G.~J., {Korista}, K.~T., {Verner}, D.~A., {Ferguson}, J.~W.,
  {Kingdon}, J.~B., \& {Verner}, E.~M. 1998, PASP, 110, 761}
\bibitem{gall+07}{{Gallagher}, S.~C., {Richards}, G.~T., {Lacy}, M., {Hines}, D.~C., {Elitzur},  M., \& {Storrie-Lombardi}, L.~J. 2007, ApJ, 661, 30}
\bibitem{hao+05}{{Hao}, L., et~al. 2005, AJ, 129, 1795}
\bibitem{hewett_foltz03}{{Hewett}, P.~C. \& {Foltz}, C.~B. 2003, AJ, 125, 1784}\bibitem{hoenig+12}{{H{\"o}nig}, S.~F., {Kishimoto}, M., {Antonucci}, R., {Marconi}, A., {Prieto},  M.~A., {Tristram}, K., \& {Weigelt}, G. 2012, ApJ, 755, 149}
\bibitem{keating+12}{{Keating}, S.~K., {Everett}, J.~E., {Gallagher}, S.~C., \& {Deo}, R.~P. 2012, ApJ, 749, 32}
\bibitem{konigl_kartje94}{{K{\"o}nigl}, A. \& {Kartje}, J.~F. 1994, ApJ, 434, 446}
\bibitem{coleman+13}{{Krawczyk}, C.~M., et.~al 2013, ApJ, submitted}
\bibitem{murray_chiang97}{{Murray}, N. \& {Chiang}, J. 1997, ApJ, 474, 91}
\bibitem{murray+95}{{Murray}, N., {Chiang}, J., {Grossman}, S.~A., \& {Voit}, G.~M. 1995, ApJ, 451, 498}
\bibitem{richards+06}{{Richards}, G.~T., et~al. 2006, ApJS, 166, 470}
\bibitem{richards+11}{--- 2011, AJ, 141, 167}
\bibitem{simpson05}{{Simpson}, C. 2005, MNRAS, 360, 565}
\bibitem{weymann+91}{{Weymann}, R.~J., {Morris}, S.~L., {Foltz}, C.~B., \& {Hewett}, P.~C. 1991, ApJ, 373, 23}
\bibitem{wolf03}{{Wolf}, S. 2003, Computer Physics Communications, 150, 99}
\end{thebibliography}
\end{document}